\documentclass[preprint,superscriptaddress,nofootinbib]{revtex4-1}

\usepackage{microtype}
\usepackage{amsmath}
\usepackage{amsfonts}
\usepackage{amssymb}
\usepackage{amsthm}
\usepackage{mathtools}
\usepackage{enumitem}
\usepackage{titlesec}
\usepackage{color}
\usepackage{graphicx}
\usepackage{amssymb}
\usepackage{hyperref}
\usepackage{yfonts}

\renewcommand{\Re}{\operatorname{Re}}
\renewcommand{\Im}{\operatorname{Im}}

\newcommand{\dd}{\mathrm{d}}
\newcommand{\ii}{\mathrm{i}}

\newcommand{\mf}[1]{\mathfrak{#1}}

\newcommand{\vph}[1]{\vphantom{#1}}

\begin{document}
	
	\title{Shear hydrodynamics, momentum relaxation, and the KSS bound}
	
	\author{Tudor Ciobanu}
	\affiliation{Department of Physics, Stanford University, \\
		Stanford, CA 94305, USA}
	\author{David M.~Ramirez}
	\affiliation{Center for Quantum Mathematics and Physics (QMAP) \\
          Department of Physics, University of California, Davis, CA 95616 USA}
	
	\begin{abstract}
          In this paper we investigate the behavior of the shear
          hydrodynamic response functions in a simple holographic
          model exhibiting momentum relaxation. We compute several
          stress tensor response functions in the transverse channel,
          and from there derive the ratio of shear viscosity to
          entropy density, $\eta/s$, using two different methods. The
          two values differ from each other, one which satisfies the
          KSS bound $\eta/s \geq \frac{1}{4\pi}$ and one which does
          not, and we discuss the causes and implications of this
          result.
	\end{abstract}
	
	\maketitle

\allowdisplaybreaks

\section{Introduction}

Over the past two decades, holography has shed new light and offered
new perspectives on hydrodynamics, interpreting decades of work on
black hole thermodynamics in terms of the behavior of the dual field
theory placed at a finite temperature. One of the more stimulating
results of holographic hydrodynamics has been the empirical
universality of the ratio of the shear viscosity to entropy density,
$\eta/s$ \cite{Policastro:2001yc,Kovtun:2003wp}. Starting with the
pioneering work of Kovtun, Son, and Starinets (KSS)
\cite{Kovtun:2004de}, the ratio has been conjectured to be bounded
from below, with the initial lower bound given by
\begin{equation}
  \label{eq:kss-bound}
  \frac{\eta}{s} \geq \frac{1}{4\pi} \frac{\hbar}{k_B}\, .
\end{equation}
While various counterexamples to the bound as written have been found
\cite{Kats:2007mq, Brigante:2008gz, Buchel:2008vz, Rebhan:2011vd,
  Mamo:2012sy, Critelli:2014kra, Ge:2014aza, Jain:2015txa,
  Hartnoll:2016tri, Alberte:2016xja, Burikham:2016roo, Ling:2016ien,
  Liu:2016njg, Wang:2016vmm, Ling:2016yxy, Chakraborty:2017msh}, the
notion of a fundamental bound on physically relevant quantities,
e.g.~hydrodynamic quantities, remains very enticing, and various
attempts have been made to either generalize the bound to incorporate
the known counterexamples \cite{Cremonini:2011iq} or otherwise extract
general lessons regarding the nature of strongly coupled dynamics
\cite{Hartnoll:2014lpa, Maldacena:2015waa, Lucas:2017ggp,
  Hartman:2017hhp}.

Physically, one way to view the ratio is to note that, up to a factor
of the temperature, it governs the diffusion of the shear momentum:
$TD_\perp = \eta/s$. Indeed, as is well known (see e.g.~\cite{forster,
  chaikin-lubensky, Kovtun:2012rj, Hartnoll:2016apf}), in the
hydrodynamic regime, the shear momentum density response function is
given by (taking the momentum $\vec{k} = k \hat{x}$ in the $x$
direction)
\begin{equation}
  \label{eq:intro-ty}
  G^R_{T^{ty},T^{ty}}(\omega,k) = {-}\frac{\eta k^2}{{-}\ii \omega + \frac{\eta}{sT} k^2}\, .
\end{equation}
By utilizing the momentum conservation Ward identity, $\eta$ can also
be extracted from the shear momentum current correlators by a Kubo
formula:
\begin{equation}
  \label{eq:intro-xy}
  \eta = {-} \lim_{\omega\to 0} \frac{\Im G^R_{T^{xy},T^{xy}}(\omega,k=0)}{\omega}\, .
\end{equation}
Thus, the shear viscosity can be alternatively viewed as a measure of
the low frequency spectral weight of the shear momentum current
density.

A common theme of recent work in holographic transport has been the
incorporation of translational symmetry breaking. Practically,
momentum relaxation naturally arises in physical systems, for example
via lattices or quenched disorder, and there has been a lot of recent
progress incorporating momentum relaxation holographically
\cite{Hartnoll:2008hs, Adams:2011rj, Hartnoll:2012rj, Adams:2012yi,
  Horowitz:2012ky, Vegh:2013sk, Blake:2013bqa, Blake:2013owa,
  Arean:2013mta, Donos:2013eha, Lucas:2014zea, Taylor:2014tka,
  Donos:2014cya, Donos:2014yya, Baggioli:2014roa, Lucas:2014sba,
  Hartnoll:2014cua, Donos:2015gia, Banks:2015wha, Hartnoll:2015faa,
  Hartnoll:2015rza, Grozdanov:2015qia, Grozdanov:2015djs,
  Lucas:2015lna, Rangamani:2015hka, Gouteraux:2016wxj,
  Fadafan:2016gmx, Blake:2016sud}. The primary theoretical motivation
for including momentum relaxation is to resolve subtleties regarding
dc transport. For example, since the heat current necessarily overlaps
with the conserved momentum operator at finite temperature,
translation invariance prevents the heat current from relaxing and
leads to an infinite dc thermal conductivity.  Holography provides a
particularly useful setting in which to explore effects of momentum
relaxation as the strongly interacting nature of holography precludes
long-lived quasiparticles which can otherwise dominate transport. The
paucity of tractable calculations involving both strong coupling and
momentum relaxation as well as the possible relevance of such settings
in many condensed matter contexts has made holography an attractive
alternative approach.

All of the holographic studies of $\eta/s$ alluded to above were in
translationally invariant settings, and therefore the shear viscosity
apparently does not suffer from subtleties related to momentum
conservation. Indeed, it would seem we have the opposite problem in
that an unambiguous definition for $\eta$ \emph{requires}
translational symmetry. The key ingredient in connecting the two
approaches to $\eta$ is the use of the structure enforced by
hydrodynamics and the resulting Ward identities. Thus, upon breaking
translation invariance, we may naturally expect that the two
prescriptions for calculating $\eta$ will no longer agree. Indeed,
without translational symmetry, momentum is no longer a good quantum
number and it's not clear how to apply the methods above to define
$\eta$. One is lead to ask which, if any, approach physically captures
the notion of shear viscosity in the presence of momentum relaxation.

Of the two interpretations discussed above, the second is well
defined. Namely the spectral weight of a particular component of the
stress tensor makes perfect sense (provided one has a stress tensor),
even if it is no longer a component of a conserved current. This
perspective and the role that $\eta/s$ plays in entropy production was
recently discussed in detail in \cite{Hartnoll:2016tri}.

On the other hand, the first interpretation is more subtle. Once
translation invariance is broken, momentum no longer diffuses and so a
corresponding `diffusion constant' is not well defined. One expects,
and we will see explicitly below, that a non-zero momentum relaxation
rate $\Gamma$ will shift the hydrodynamic pole into the lower half
plane. However, for a sufficiently weak momentum relaxation rate
$\Gamma \ll T$, an approximate diffusion constant can be defined for
length scales $\ell \ll \Gamma^{{-}1}$, as long as the length scale is
large enough such that hydrodynamics can be trusted. Practically, one
can simply search for the pole in the shear momentum density
correlators and extract a diffusion constant from the momentum
dependence of the dispersion relation. However, the relation, if any,
of this quantity to the $\eta$ obtained by the Kubo formula is not
obvious \cite{Hartnoll:2016apf}. The interplay between shear momentum
transport and momentum relaxation, particularly in the presence of
spontaneous translational symmetry breaking, has also recently been
discussed in \cite{Delacretaz:2017zxd}.

In this brief note, we will investigate the behavior of the shear
hydrodynamic response functions in a simple holographic model
incorporating momentum relaxation. Our primary objective is to compare
the holographic results to simple hydrodynamic arguments and also
compare the notions of the shear viscosity discussed above. In
particular, our key result is that the value of $\eta/s$ differs
depending on whether it is obtained via the Kubo formula or via the
pseudo-diffusive pole in the shear momentum density two-point
function. This is not entirely surprising as the equivalence of these
two approaches in traditional hydrodynamics relies on momentum
conservation, which we are manifestly breaking. Explicitly, we can see
that the Kubo formula calculates the spectral weight strictly at
$k=0$, and for any small but fixed momentum relaxation rate the
momentum will have decayed at these length scales. Interestingly, the
value obtained via the pseudo-diffusive pole satisfies the KSS bound
\eqref{eq:kss-bound}, while the Kubo formula yields a value of
$\eta/s< \frac{1}{4\pi}$ as shown in \cite{Hartnoll:2016apf}. 

The rest of the paper is organized as follows. In
Sec.~\ref{sec:hydro}, we study the effect of momentum relaxation in a
simple hydrodynamic toy model, obtaining predictions for the retarded
shear correlation functions $G_{T^{ty}T^{ty}}(\omega,k)$ and
$G_{T^{xy}T^{xy}}(\omega,k)$. Throughout the paper we focus on $2+1$
dimensional systems for concreteness. In Sec.~\ref{sec:axion-thy}, we
introduce the holographic model which will be our focus for the
remainder of the paper and review its known thermodynamic and
transport properties. In the remaining two sections,
Sec.~\ref{sec:ty-corr} and Sec.~\ref{sec:xy-corr}, we evaluate the
shear correlators holographically, working perturbatively in the
momentum relaxation rate, and these results are compared to the
predictions by our hydrodynamic toy model. Various technical details
of the calculations are relegated to appendices.

\section{Shear hydrodynamics and momentum relaxation}
\label{sec:hydro}
Before turning to our holographic analysis, we investigate a simple
mechanism for momentum relaxation in the context of relativistic
hydrodynamics. In this scenario, standard arguments due to Kadanoff
and Martin \cite{Kadanoff1963} allow for evaluation of the retarded
two-point functions of interest. For modern discussions of the
approach, see for example \cite{Kovtun:2012rj, Davison:2014lua,
  Hartnoll:2016apf}.

For simplicity, we focus on neutral, conformal fluids. The hydrodynamic
equations of motion, i.e.~the energy-momentum conservation equations,
read
\begin{equation}
  \label{eq:hydroeom}
  \partial_\mu T^{\mu \nu} = 0\, .
\end{equation}
These equations, when supplemented with the appropriate constitutive
relations, govern the relaxation of energy and momentum
fluctuations. We concentrate on the shear channel, taking the
fluctuations to be plane waves with momentum along the $x$ direction.
More explicitly, we turn on a small source for the transverse momentum
density $P^y \equiv T^{ty}$, namely a velocity fluctuation $\delta
v^y(t,x)$. To lowest order in $\omega$ and $k$, linear response and
the constitutive relations express the fluctuations of the momentum
density, $\delta P^y$, and momentum current, $\delta T^{xy}$, in terms
of $\delta v^y$ as:
\begin{align}
  \label{eq:const-rel}
  \delta P^y(\omega,k) ={}& \chi_{PP} \delta v^y(\omega,k) = (\epsilon+P) \delta
  v^y(\omega,k)\, , & \delta T^{xy}(\omega,k) ={}& {-} \ii \eta k \delta v^y
  (\omega,k) \, .
\end{align}
Here $\chi_{PP} = \epsilon+P = sT$ is the static susceptibility
relating the source and response fluctuations, and $\epsilon$, $P$,
$\eta$, $s$ are the energy density, pressure, shear viscosity, and
entropy density respectively. Using these relations in the (Fourier
transformed) momentum conservation equation $\partial_t \delta P^y +
\ii k \delta T^{xy} =0$ yields a diffusion equation for the
fluctuation $\delta P^y$ with diffusion constant $D_\perp =
\frac{\eta}{\epsilon + P} = \frac{\eta}{sT}$.

The analysis so far assumes momentum conservation. However, we are
interested in the consequences of momentum relaxation. For simplicity,
we assume that this perturbation does not change the form of the
constitutive relation and that to lowest order its only effect is to
modify the conservation equation. A simple ansatz for this effect is
to simply add a constant momentum relaxation rate to the conservation
equation
\begin{equation}
  \label{eq:mom-cons-gamma}
  \partial_t \delta P^y + \ii k \delta T^{xy} = {-} \Gamma \delta P^y\, .
\end{equation}
Using the constitutive relations \eqref{eq:const-rel}, we can rewrite
this result as
\begin{equation}
  \label{eq:dtPy-deltav}
  \partial_t \delta P^y + \left[\Gamma (\epsilon+P) + \eta k^2\right] \delta v^y = 0\, .
\end{equation}
Appealing now to standard results of linear response
\cite{forster,Kovtun:2012rj,Hartnoll:2016apf}, the momentum density
retarded Green's function is given by
\begin{align}
  \label{eq:hydro-py-corr}
  G^R_{P^y,P^y}(\omega,k) ={}& {-}(\epsilon+P) \frac{\Gamma +
    \frac{\eta}{\epsilon+P} k^2}{{-}\ii \omega + \Gamma +
    \frac{\eta}{\epsilon+P} k^2}\, .
\end{align}
As expected, the diffusive pole in the shear momentum correlator has
been pushed further into the lower half of the complex plane by a
constant amount set by the momentum relaxation rate $\Gamma$.

Unfortunately, there is no analogous approach to determine
$G^R_{T^{xy},T^{xy}}$ in the presence of momentum relaxation. When
$\Gamma=0$, momentum conservation implies corresponding Ward
identities, which must hold as operator identities and allow us to
determine $G^R_{T^{xy},T^{xy}}$ from $G^R_{T^{ty},T^{ty}}$ (up to
contact terms). However, our modified conservation equation
\eqref{eq:mom-cons-gamma} is not such an operator identity, at least
without more knowledge about the mechanism of momentum relaxation, and
therefore we cannot use the same techniques. Furthermore, without a
dynamical equation of motion for $T^{xy}$, which would go beyond our
hydrodynamic considerations here, direct application of the
Kadanoff-Martin approach is not possible.

\section{Neutral linear axion model}
\label{sec:axion-thy}

In this section we introduce the holographic model that will be our
focus for the remainder of this note. These models, known as linear
axion models (among other names), were introduced in
\cite{Andrade:2013gsa} as simple examples in which translational
symmetry on the boundary can be broken while retaining a homogeneous
metric, and they have been served as useful models to investigate the
effects of momentum relaxation in recent years \cite{Alberte:2015isw,
  Ge:2014aza, Alberte:2016xja, Hartnoll:2016apf,
  Caldarelli:2016nni}. The homogeneity of the metric reduces the bulk
equations of motion reduce to ODEs rather than PDEs, dramatically
simplifying calculations.

The model consists of two massless scalars, $\phi^i$ with $i=1,2$,
minimally coupled to gravity, with the action given by
\begin{equation}
  \label{eq:action}
  S = \frac{1}{2\kappa^2} \int \dd^4 x\, \sqrt{-g} \left[R + \frac{6}{L^2} - \frac{1}{2} \sum_i (\partial_a \phi^i) (\partial^a \phi^i) \right] + S_{GH} + S_{ct}\, .
\end{equation}
Here $\kappa^2 = 8 \pi G_N$ is the gravitational coupling constant and
$L$ is the AdS radius (which we will set to $1$ throughout). This bulk
action has to be supplemented with additional terms that depend only
on the boundary data, namely the Gibbons-Hawking boundary term
$S_{GH}$ and a counterterm action $S_{ct}$. These terms, which are
only needed for the analysis of the on-shell action are given in
\autoref{ap.3}. The equations of motion following from the bulk action read
\begin{align}
  R_{ab} + 3 g_{ab} ={}& \frac{1}{2} \sum_i \partial_a
  \phi^i \partial_b \phi^i\, , \label{eq:EE} \\
  \Box \phi^i ={}& 0\, . \label{eq:KG-eq}
\end{align}

A simple class of solutions to the bulk equations of motion can be
found with the following form
\begin{align}
  \label{eq:metric-bg}
  \dd s^2 ={}& \frac{1}{r^2} \left[{-}f(r) \dd t^2 + \frac{\dd
      r^2}{f(r)} +\dd x^2 + \dd y^2 \right]\, ,  \\
  \label{eq:scalar-bg}
  \phi^i ={}& m x^i\, , \\
  \label{eq:f-bg}
  f(r) ={}& 1 - \frac{m^2}{2}r^2 - \left(1 -\frac{m^2 r_+^2}{2}
  \right) \left(\frac{r}{r_+}\right)^3\, .
\end{align}
We'll often rescale the radial coordinate to $u= r/r_+$, so the
boundary and horizon are at $u=0$ and $u=1$ respectively. Since the
$\phi^i$ are massless scalar fields, the holographic dictionary tells
us that they correspond to sources for marginal operators on the
boundary ($\Delta = d+1 = 3$). Therefore this background describes a
scenario where a conformal fluid at finite temperature is deformed by
two marginal operators with sources linear in the spatial
directions. These sources thus explicitly break the boundary
translational symmetry, where the strength of the symmetry breaking is
characterized by $m$.

The conformal boundary is located at $r\to 0$ ($u\to 0$), while there
is a non-degenerate horizon for $m^2r_+^2 < 6$. The horizon radius
determines the thermodynamic properties via
\begin{align}
  \label{eq:axion-thermo}
  4 \pi T ={}& \frac{1}{r_+} \left( 3 - \frac{m^2 r_+^2}{2} \right)\,
  , & s ={}& \frac{1}{4 G_N r_+^2} = \frac{32\pi^3 T^2}{9 \kappa^2}\,
  .
\end{align}

As mentioned above, this class of backgrounds have proven useful in
studies of the role of momentum relaxation in transport behavior. For
small $m$, i.e.~$m\ll T$, the momentum is almost conserved and this
long-lived quantity will essentially determine the transport. This
scenario falls under the name of coherent transport. However, the
background specified by \eqref{eq:metric-bg}--\eqref{eq:f-bg} exists
for any $m^2 r_+^2 < 6$, and so by dialing the parameter $m$, we can
tune from coherent transport to incoherent transport. For a detailed
analysis explicitly demonstrating the coherent-to-incoherent crossover
in the thermal transport of these models, see \cite{Davison:2014lua}.

As discussed in \cite{Vegh:2013sk, Blake:2013owa, Davison:2013jba},
the breaking of translational symmetry on the boundary is manifested
in the bulk as giving a mass to the graviton. In particular,
\cite{Hartnoll:2016tri} emphasized how a non-zero graviton mass
generically leads to a violation of the KSS bound on the shear
viscosity to entropy density ratio. For the linear axion models, the
KSS ratio was shown in \cite{Hartnoll:2016tri} to be given,
perturbatively in $m/T$, by
\begin{equation}
  \label{eq:KSS-ratio}
  4\pi\frac{\eta}{s} = 1 + \frac{\sqrt{3}}{16\pi} \left(1 - \frac{3\sqrt{3} \log 3}{\pi} \right) \left( \frac{m}{T} \right)^2 + {\cal O}\left[ \left(\frac{m}{T} \right)^4 \right]\, .
\end{equation}
As discussed in the introduction, while an obvious hydrodynamic
interpretation of the shear viscosity breaks down in the absence of
translational symmetry, it was argued that the shear viscosity, as
defined by the Kubo formula, retains a fundamental interpretation as
determining the low energy spectral weight and hence the rate of
entropy production when subjected to a slowly varying strain $\delta
g_{xy}^{(0)}$.

In the following sections, we will move away from the strict $\omega =
0$ transport behavior and study the low $\omega$, $k$ shear
correlation functions perturbatively in $m$. The structure of these
correlators is determined, at $m=0$, by hydrodynamics, and in
particular momentum conservation, for sufficiently small $\omega,
k$. Therefore, we expect that there will be non-trivial interplay
between the $m \to 0$ and $\omega \to 0$ limits, since for any finite
$m$, the momentum is no longer conserved and therefore decays at late
times.

\section{Calculating correlators}

In this section we will apply standard holographic techniques to
determine the correlator $G_{T^{ty},T^{ty}}^R$ in the backgrounds
discussed in the previous section \cite{Policastro:2002se,
  Iqbal:2008by, Hartnoll:2009sz, McGreevy:2009xe,
  Hartnoll:2016apf}. The basic strategy is to perturb our background
solutions, which corresponds to turning on a small source on the
boundary, solve the linearized bulk equations of motion with
appropriate boundary conditions at the conformal boundary and ingoing
boundary conditions at the horizon, and extract the response of the
system from the near boundary behavior. Since we are interested in
correlators of the shear stress tensor components $T^{ty}$ and
$T^{xy}$, we will focus on the behavior of the shear metric
fluctuations $\delta g_{\mu\nu} = h_{\mu \nu}$, which are coupled via
the equations of motion to the scalar fluctuations $\delta \phi^2$.

Exploiting the homogeneity of our background solution, we Fourier
decompose our fluctuations and take them to have the form $\delta X =
X(u) e^{{-}\ii (\omega t - k x)}$, where $X \in \{h_{ty}, h_{xy},
\delta \phi^2\}$ and we've also used isotropy to set the momentum
along the $x$ direction without loss of generality. The linearized
equations of motion for the shear modes then read
\begin{align}
  0&=\frac{u^2}{r_+^2} \frac{\dd}{\dd u} \left(\frac{f
      h_x^{y\prime}}{u^2}\right) + \frac{\omega}{f}
  (\omega h_x^y+kh_t^y) -m^2 h_x^y+\ii km\delta \phi_2 \label{eq:hxy-eom} \, ,\\
  0&= \frac{u^2}{r_+^2} \frac{\dd}{\dd u} \left(\frac{f
      h_t^{y\prime}}{u^2}\right) - \frac{k}{f} (\omega
  h_x^y+kh_t^y)-\frac{m^2}{f}h_t^y-\frac{\ii \omega m}{f} \delta
  \phi_2 \label{eq:hty-eom} \, ,  \\
  0&= \frac{u^2}{r_+^2}\frac{\dd}{\dd u} \left(\frac{f \delta \phi_2
      '}{u^2}\right) +\frac{1}{f} (\omega^2-k^2f)\delta \phi_2 -
  \frac{\ii m}{f} (\omega
  h_t^y+kfh_x^y) \label{eq:sc-eom} \, , \\
  0&= \ii\omega h_t^{y\prime} +\ii k f h_x^{y\prime} -mf \delta \phi_2
  ' \, .  \label{eq:fluc-constr}
\end{align}
Here the primes denote $u$ derivatives, indices are raised using the
background solution from the previous section, and for simplicity,
we've used the gauge freedom to set $h_r^y=0$.

The strategy for extracting the low $\omega$, $k$ behavior will be as
follows. We impose the infalling boundary
conditions at the horizon by writing the radial profile of the
fluctuations as $X(u) = f(u)^{-\ii \omega/4\pi T} \tilde
X(u)$. Plugging this form into the wave equations above we obtain a
new set of equations for $\tilde X$, which we will then solve
perturbatively in $\omega$, $k$, and $m$. Once we have these
solutions, we can examine their near boundary behavior to extract the
Green's functions of interest.

\subsection{Solving for $G_{T^{ty}, T^{ty}}$}
\label{sec:ty-corr}
We first look at $G^R_{T^{ty}, T^{ty}}$. This correlator is determined
by the mode $h_{ty}$, so we want to decouple $h_{ty}$ from $h_{xy}$
and $\delta \phi^2$. This is easily accomplished by differentiating
\eqref{eq:hty-eom} and combining with \eqref{eq:fluc-constr}, which
yields a third order ODE for $h_{ty}$. If we define $\psi_{ty} =
-\frac{1}{r_+^2 u} h^{y\prime}_t$, we can write this equation as
\begin{equation}
  \label{eq:psi-ty-eq}
  0 = \frac{\dd}{\dd u} (f \psi_{ty}') + \frac{r_+^2 \omega^2 - r_+^2 k^2 f - r_+^2 m^2f - \frac{1}{u^2} f f'}{f} \psi_{ty}\, .
\end{equation}

In order to bring the equation to a form easier to solve we will 
introduce the dimensionless variables
\begin{align}\label{eq:parameters}
  w :={}& \frac{\omega}{2\pi T}\, , & q :={}& \frac{k}{2\pi
    T}\, , & M:={}& \frac{m}{2\pi T}\, , & R_+
  :={}& 2\pi T r_+ = \frac{3}{2}-\frac{M^2R_+^2}{4} \, .
\end{align}
Note that since $R_+$ is defined implicitly, to work perturbatively in
$M$, we will expand $R_+ = \frac{3}{2} - \frac{9M^2}{16} + {\cal
  O}(M^4)$.

With the new dimensionless variables, the ODE and emblackening factor reads
\begin{align}
  0 ={}& \psi''_{ty} + \frac{f'}{f} \psi'_{ty} +
  \frac{R_+^2(w^2-q^2f-M^2f)-\frac{1}{u} ff'}{f^2} \psi_{ty}\, , \\
  f(u) ={}& 1 - \frac{1}{2} M^2 R_+^2 u^2 - \left( 1 -
    \frac{M^2 R_+^2}{2} \right) u^3 \, .
\end{align}

As described above, in order to deal with the singular point of the
ODE we scale out the singular power law behavior at the horizon $u \to
1$, writing $\psi_{ty}(u) = f(u)^{{-} \ii w/2} \tilde
\psi_{ty}(u)$ to obtain the following ODE for $\tilde \psi_{ty}$
\begin{equation}
  \label{eq:psi-ty-tilde-eq}
  0 = \tilde \psi_{ty}'' + (1 - \ii w) \frac{f'}{f} \tilde
  \psi_{ty}' + \left[-\frac{1}{u} \frac{f'}{f} - \ii \frac{w}{2}
    \frac{f''}{f}+ \frac{w^2}{f^2} \left(R_+^2 -
      \frac{f'^2}{4}\right) - \frac{R_+^2(q^2+M^2)}{f}
  \right] \tilde \psi_{ty}\, .
\end{equation}
We are now ready to solve this equation perturbatively. To do so, we
simply expand
\begin{align}
  \label{eq:psi-ty-pert}
  \tilde \psi_{ty} = \tilde \psi_{ty}^{(0)} + \ii w\tilde
  \psi_{ty}^{(1)} + q^2 \tilde \psi_{ty}^{(2)} + M^2 \tilde
  \psi_{ty}^{(3)} + \ii w M^2\tilde \psi_{ty}^{(4)}+ q^2 M^2 \tilde
  \psi_{ty}^{(5)} + \dotsb\, .
\end{align}
Plugging this into \eqref{eq:psi-ty-tilde-eq}, we can solve order by
order in $w$, $q$, and $M$. Note that the factors of $R_+$, including
those in the emblackening factor, depend on $M$, so for consistency
one needs to expand it appropriately as well.

Once we have the solution, we want to expand near the boundary $u=0$
to extract the Green's function. By expanding \eqref{eq:psi-ty-eq} near
$u=0$, or using the known asymptotics of $h_{\mu\nu}$ and tracing
through the definition of $\psi_{ty}$, we see that near the boundary
$\psi_{ty}$ must take the schematic form $\psi_{ty} \sim {\cal A} +
{\cal B} u$. More explicitly, if we write the near boundary metric
perturbations as
\begin{equation}
  \label{eq:hty-near-bdy}
  h^y_t = h^{y(0)}_t + u h^{y(1)}_t + u^2 h^{y(2)}_t + u^3 h^{y(3)}_t + \dotsb \, ,
\end{equation}
we can use the linearized equations of motion to find $h^{y(1)}_t = 0$
and $h^{y(2)}_t = r_+^2\frac{k^2 + m^2}{2} h^{y(0)}_t$, which upon
plugging into our definition of $\psi_{ty}$ tells us that ${\cal A} =
2 h^{y(2)}_t$ and ${\cal B} = 3h^{y(3)}_t$. Since $h^{y(3)}_t$ is the
first variation of the on-shell action with respect to the source
$h^{y(0)}_t$ (see \autoref{ap.3}), the definition of the retarded
Green's function via linear response indicates
\begin{align}
  \label{eq:Gtyty-AB}
  G_{T^{ty},T^{ty}}^R(\omega,k) = \frac{3}{2r_+^3\kappa^2}
  \frac{h^{y(3)}_t}{h^{y(0)}_t} = \frac{k^2+m^2}{2r_+ \kappa^2}
  \frac{{\cal B}}{{\cal A}}\, .
\end{align}
Here the factor of $r_+^3$ comes from expanding near the boundary in
$u$ instead of $r$.

To spare the reader, we've collected the details of obtaining the
solution in \autoref{ap.1}, and we'll only quote the final result,
which reads
\begin{align}
  \label{eq:psi-tysol}
  C^{-1}\psi_{ty}(u) ={}& u +\frac{3}{4}\left(1-u\right)\left(q^2+M^2\right) \\
  &- \frac{1}{12} \ii w \left\{ 18 + u \left[6\sqrt{3}
      \tan^{-1}\left(\frac{1+2u}{\sqrt{3}} \right) + 9 \log \left(
        \frac{3}{1+u+u^2} \right) -18 - 2 \sqrt{3} \right] \right\}
  \nonumber \\
  &+ \dotsb\, . \nonumber
\end{align}
Here $C$ is an integration constant that sets the scale of the source
we've turned on. Expanding this near the boundary we find
$\mathcal{A}$ and $\mathcal{B}$ are given by
\begin{align}
  \mathcal{A} ={}& C \left[ -\frac{3}{2} \ii w + \frac{3}{4} \left(k^2 +
    M^2\right)
    + \dotsb \right]\, ,  \\
  \mathcal{B} ={}& C \left[ 1 + \frac{\ii w}{12} \left( 18 + \sqrt{3}
      \pi - 9 \log 3 \right) - \frac{3}{4}\left(q^2+M^2\right) \right] \, .
\end{align}
From the expression for ${\cal A}$, we can easily find the location of
the poles in $G_{T^{ty}, T^{ty}}^R$:
\begin{equation}
  w= -\ii \frac{1}{2} \left(q^2+M^2\right)\, ,
\end{equation}
or, reinstating the factors of $T$ from \eqref{eq:parameters} we get
that
\begin{align}
  \omega = - \ii \left[\frac{k^2}{4\pi T} + \frac{m^2}{4\pi
      T}\right]\, . \label{eq:ty-pole-0}
\end{align}
Here we see that, as expected, the diffusive pole gets pushed into the
lower half plane by the momentum relaxation.

All in all, we can write the shear momentum correlator as
\begin{equation}
  G_{T^{ty}T^{ty}}^R(\omega,k) =  \frac{1}{2\kappa^2} \frac{16 \pi^2 T^2}{9} \frac{k^2+m^2}{- \ii \omega + \frac{k^2+m^2}{4 \pi T}}\, .  
\end{equation}
As an aside, we note that with this expression for $G_{T^{ty},
  T^{ty}}^R$, we can readily obtain the the optical thermal
conductivity via the Kubo formula\footnote{Here we denote that thermal
  conductivity by $\kappa_T$ in order to differentiate it from the
  gravitational coupling $\kappa^2$.}
\begin{align}
  \kappa_T(\omega) ={}& \frac{1}{T} \frac{G_{T^{ty}T^{ty}}
    (\omega,k=0) -G_{T^{ty}T^{ty}} (0,k=0)}{\ii \omega } = \frac{32}{9
    \kappa^2}\frac{\pi^3 T^2}{-\ii \omega + \frac{m^2}{4\pi T}} =
  \frac{s}{{-}\ii \omega + \frac{m^2}{4\pi T}}\,
  .
\end{align} 
For the last equality, we've used our expression for the entropy
density \eqref{eq:axion-thermo}. Here we see explicitly that the
momentum relaxation rate resolves the divergence in the DC thermal
conductivity, leading to a finite Drude peak with width ${\cal
  O}(m^2/T)$, as discussed extensively in \cite{Davison:2014lua}.

In the solution above, we see that the ``diffusion'' constant, that is
the coefficient of $k^2$ in the pole \eqref{eq:ty-pole-0}, is
unchanged from the AdS$_4$ value, $\frac{\eta}{sT} = \frac{1}{4\pi
  T}$. To see how the momentum relaxation affects this ``diffusion'',
one needs to continue to higher orders, including terms of order $M^2
w$ and $M^2 q^2$. It is in fact possible to solve this equation
non-perturbatively in $M$, to lowest order in $w$ and $q$. That is,
taking $\tilde \psi_{ty} = \tilde \psi_{ty}^{(0)}(u;M) + w\tilde
\psi_{ty}^{(1)}(u;M) + q^2 \tilde \psi_{ty}^{(2)}(u;M)$, and
substituting into \eqref{eq:psi-ty-tilde-eq}, one can solve for
$\tilde \psi_{ty}^{(0)}(u;M)$, $\tilde \psi_{ty}^{(1)}(u;M)$, and
$\tilde \psi_{ty}^{(2)}(u;M)$ with generic $M$. As the expressions are
rather unwieldy, we won't present this full solution, though we
present the necessary equations in \autoref{ap.1}. The
``hydrodynamic'' pole is obtained by finding the zero in ${\cal A}$,
and working to the next non-trivial order in $M^2$, one finds the
following for ${\cal A}$
\begin{align}
  C^{-1} {\cal A} ={}& -\frac{3}{2} \left[1 - \frac{1}{24} \left(9 \log
      3 - 9 - \sqrt{3}\pi \right) M^2 \right] \ii w +
  \frac{3}{4} q^2 + \frac{3}{4} \left(1 - \frac{3}{8}M^2 \right) M^2 +
  \dotsb\, , \label{eq:ty-A-m2}
\end{align}
which gives a pole at
\begin{align}
  \ii w = \frac{q^2}{2} \left[1 + \frac{1}{24} \left( 9 + \sqrt{3}\pi
      - 9 \log 3 \right) M^2 \right] + \frac{M^2}{2} + {\cal O}(M^4)\, .
\end{align}
Therefore, upon restoring factors of $T$, we see that the
``diffusion'' constant is given by
\begin{align}
  \label{eq:TD-pole}
  D_\perp = \frac{1}{4\pi T} \left[1 + \frac{1}{24} \left( 9 + \sqrt{3}\pi -
      9 \log 3 \right) M^2 \right] + \dotsb\, .
\end{align}
Comparing this with the result found in \cite{Hartnoll:2016tri},
reproduced in \eqref{eq:KSS-ratio}, we see that the ratio $4\pi
\eta/s$ obtained from the ``diffusion'' constant of the pole in
$G_{T^{ty},T^{ty}}^R$ does not agree with the value found via the Kubo
formula. However, it is curious to note that the value
\eqref{eq:TD-pole} satisfies the KSS bound \eqref{eq:kss-bound}.

\subsection{Calculating $G_{T^{xy}T^{xy}}$}
\label{sec:xy-corr}

We now turn to the correlator $G^R_{T^{xy},T^{xy}}$. Unlike the
$T^{ty}$ case, here it won't be possible to fully decouple the metric
and scalar fluctuations for $k\neq 0$. We'll start by considering the
simpler $k=0$ case, which in any case will allow us to rederive the
results for the shear viscosity in \cite{Hartnoll:2016tri}. 

\subsubsection{$k=0$}
At $k=0$, the $h^y_x$ equation of motion completely decouples $h^y_x$
from the other modes, as can be easily read off from
\eqref{eq:hxy-eom}. With a little rearranging, the $k=0$ equation
reads
\begin{equation}
  \label{eq:txy-hxy-eq-k0}
  0 = \frac{\dd}{\dd u} \left( \frac{f h^{y\prime}_x}{u^2} \right) + \frac{r_+^2(\omega^2 - m^2 f)}{u^2 f} h^y_x\, .
\end{equation}
Again, a quick check of the on-shell action (see \autoref{ap.3})
confirms that the Green's function can be read off of the solution
(satisfying infalling boundary conditions at the horizon) to this
equation via
\begin{equation}
  \label{eq:GR-xy-bdy-data}
  G_{T^{xy}T^{xy}}^R =  \frac{3}{2 r_+^3  \kappa^2} \frac{h^{y(3)}_x}{h^{y(0)}_x} = \frac{3(m^2-\omega^2)}{2 r_+ \kappa^2} \frac{h^{y(3)}_x}{h^{y(2)}_x}  \, .
\end{equation}
Here we've used the equations of motion to relate the subleading
behavior $h^{y(2)}_x$ to the leading behavior $h^{y(0)}_x$, simply as
a technical convenience. With the two point function in hand, we can
compute the shear viscosity and modulus via:
\begin{align}
  \label{eq:txy-kubo}
  \eta ={}& {-} \lim_{\omega \to 0} \frac{\Im
    G^R_{T^{xy},T^{xy}}(\omega,k=0)}{\omega}\, , & {\cal G} ={}&
  \lim_{\omega \to 0} \Re G_{T^{xy}T^{xy} }(\omega,
      k=0)\, .
\end{align}

To explicitly determine the two point function, we solve
\eqref{eq:txy-hxy-eq-k0} perturbatively as in the previous
section. The details are relegated to the appendix, but to orient
ourself we note that the object we'll need is the ratio $\mathcal{R}=
\frac{h^{y(2)}_x}{h^{y(3)}_x}$, which we will expand for small $\omega$ as
\begin{equation}
  {\cal R} = \mathfrak{a} + \ii \mathfrak{b} +
  \mathfrak{e}w + \ii \mathfrak{f} w + \mathfrak{g} w^2 + \ii \mathfrak{h}
  w^2 + \mathcal{O}( w^3)\, .
\end{equation}
where we $\mathfrak{a},\dots,\mathfrak{h}$ are real, $m$-dependent
constants. Note in particular that we are taking the small $\omega$
limit first. Time reversal symmetry implies that $\Im
G_{T^{xy},T^{xy}}(\omega)$ is odd in $\omega$, and so we must take
$\mf{b}=0$ (as can be explicitly checked). Therefore we can expand the
two point function as
\begin{align}
  G^R_{T^{xy}, T^{xy}}(\omega,k=0) = \frac{3 m^2}{2 \kappa^2 \mf{a}}
  \left[1 - \frac{\mf{e} +\ii \mf{f}}{\mf{a}} \frac{\omega}{2\pi T}
  + {\cal O}(\omega^2)\right]\, .
\end{align}
Using this result and \eqref{eq:txy-kubo}, we can write the shear
viscosity and modulus as
\begin{align} \label{eq:eta-G-mf} 
  \eta ={}& \frac{3m^2}{4\pi T \kappa^2 } \frac{\mf{f}}{\mf{a}^2}\, , & 
  {\cal G} ={}& \frac{3m^2}{2\kappa \mf{a}}\, .
\end{align}
As we'll see shortly, $\mf{a} \sim {\cal O}(1)$ while $\mf{f} \sim
m^{{-2}}$ for small $m$, and so the $m\to 0$ limit reproduces the
expected results for the planar AdS black brane.

Now let's solve \eqref{eq:txy-hxy-eq-k0} for $h_x^y(u)$ exactly as we did
for $\psi_{ty}$ in the previous section. Namely, we can take the
ansatz $h_x^y(u) = f(u)^{-\ii \frac{w}{2}} F(u)$, to impose infalling
boundary conditions, and expand the remaining factor in $w$ and $x$
\begin{equation}\label{eq:psi-xy-pert}
F(u)=F_0(u)+wF_1(u) + \sum_{n=1}^{N} M^{2n} H_n(u) + \sum_{n=1}^{N} wM^{2n} J_n(u) + \dots.
\end{equation}
The explicit solutions for $F_0(u), F_1(u),$ etc.~can be found in
\autoref{ap.2}. Expanding these near $u=0$ we can find $h^{y(3)}_x$
and $h^{y(2)}_x$ and thus $G^R_{T^{xy},T^{xy}}$. 

Using the solutions in \autoref{ap.2}, we find the small $x$ behavior
\begin{align}
  \mathfrak{a} ={}& {-} \frac{3}{2} + \frac{1}{4} (\sqrt{3} \pi - 3
  \log{3}) M^2 + \mathcal{O} (M^3)\, , \\
  \mathfrak{f} ={}& \frac{9}{4x^2} - \frac{1}{8}(3 + 3\sqrt{3} \pi) +
  \mathcal{O}(x^2)\, .
\end{align}
Putting together the results, we obtain:
\begin{equation}
4 \pi \frac{\eta}{s} = 1+ \left(\frac{\pi}{3\sqrt{3}} - \log{3}\right) \left(\frac{3m}{4\pi T}\right)^2 + \mathcal{O} (x^3)\, ,
\end{equation}
which agrees with the shear viscosity obtained in
\cite{Hartnoll:2016tri}, and
\begin{equation}
  \mathcal{G} = - \frac{4\pi T}{3}m^2 - \frac{1+\pi \sqrt{3} - 3\log{3}}{8\pi T}m^4 + \mathcal{O}(m^5)\, .
\end{equation}
Here we see that the shear viscosity as computed by the Kubo formula
disagrees with the value obtained from $G^R_{T^{ty},T^{ty}}$ and that
the shear modulus is non-trivial in these backgrounds. This result on
the shear modulus has been previously emphasized in these models in
\cite{Alberte:2015isw, Alberte:2016xja}.

\subsubsection{$k \neq 0$}
Now we relax the condition $k=0$. By taking appropriate combinations
of \eqref{eq:hxy-eom} and \eqref{eq:fluc-constr}, we obtain the
following equation for $\psi_{xy} = \frac{f}{u} h_x^{y \prime}$
\begin{align}
  0 ={}& \partial_u (f \partial_u \psi_{xy}) - R_+^2\left(M^2 + q^2 -
    \frac{w^2}{f} + \frac{f'}{R_+^2 u} \right) \psi_{xy} - \frac{M^2
    R_+^2 f'}{u} \varphi\, ,
\end{align}
where $\varphi = h_x^y - \ii \frac{q}{M} \delta \phi_2$. We see that
the scalar fluctuation doesn't fully decouple from the $h_x^y$
fluctuation, and so we also need to solve for $\varphi$, using its
equation of motion
\begin{align}
  0 ={}& u^2 \partial_u \left( \frac{f \partial_u\varphi}{u^2} \right)
  - R_+^2 \left(M^2 + q^2 - \frac{w^2}{f} \right) \varphi\, .
\end{align}

To extract the Green's function from $\psi_{xy}$, we proceed as we did
for the $ty$ fluctuations and note that its near boundary expansion is
given by
\begin{align}
  \psi_{xy} = 2 h_x^{y(2)} + 3 h_x^{y(3)} u + \dotsb\, ,
\end{align}
and so using $2h_x^{y(2)} = r_+^2(\omega^2 - m^2) h_x^{y(0)}$, we have
\begin{align}
  G^R_{T^{xy},T^{xy}} = \frac{3}{2r_+^3\kappa^2}
  \frac{h_x^{y(3)}}{h_x^{y(0)}} = \frac{\omega^2 - m^2}{2 r_+
    \kappa^2} \frac{{\cal B}}{{\cal A}}\, ,
\end{align}
where ${\cal B} = \psi_{xy}'(0) = 3 h_x^{y(3)}$ and ${\cal A} =
\psi_{xy}(0) = 2 h_x^{y(2)}$.

To impose the infalling boundary conditions, we write $\psi_{xy} (u) =
f^{-i\frac{w}{2}}(u) \tilde\psi_{xy}(u)$ and $\varphi(u)=
f^{-i\frac{w}{2}} \tilde\varphi(u)$ and expand as:
\begin{align}
  \tilde \psi_{xy}(u)& = \tilde \psi_{xy}^{(0)}(u) + w\tilde
  \psi_{xy}^{(1)}(u) + q^2 \tilde \psi_{xy}^{(2)}(u) +M^2 \tilde
  \psi_{xy}^{(3)}(u) +   \dotsb\, , \\
  \tilde \varphi(u)& = \tilde \varphi^{(0)}(u) + w\tilde
  \varphi^{(1)}(u) + q^2 \tilde \varphi^{(2)}(u) + M^2 \tilde
  \varphi^{(3)}(u) + \dotsb\, . 
\end{align}
The equations governing $\tilde \psi_{xy}^{(0)}$, $\tilde
\psi_{xy}^{(1)}$, and $\tilde \psi_{xy}^{(2)}$ are the same as the
corresponding ones for $\tilde \psi_{ty}$ (as they must by symmetry),
and one can easily find $\tilde\varphi^{(0)} =C_1$ (see
\autoref{ap.2}). The only ODE which is changed is the one for $\tilde
\psi_{xy}^{(3)}$:
\begin{equation}
  0 = \left[(1-u^3)\partial_u^2 - 3u\left(u \partial_u - 1 \right) \right] \tilde \psi_{xy}^{(3)} - \frac{9}{4} \left( \tilde \psi_{xy}^{(0)}+ 3 u \tilde \varphi^{(0)} \right)\, .
\end{equation}
The solution to the above equation, requiring that non-analytic
terms vanish and that $\tilde \psi_{xy}^{(3)}(u=1) = 0$, is :
\begin{equation}
  \tilde \psi_{xy}(u) = \frac{3}{4} (C_0 - 3 C_1) \, ,
\end{equation}
where the constants $C_0$ and $C_1$ are determined by $\tilde
\psi_{xy}^{(0)} = C_0 u$ and $\tilde \varphi^{(0)} = C_1$. 

Combining this with the expressions for $\tilde \psi_{xy}^{(i)} =
\tilde \psi_{ty}^{(i)}$ (for $i<3$) from \autoref{ap.2}, we can
readily expand $\psi_{xy}$ near the boundary to find an expression of
the form $\psi_{xy}(u) = {\cal A} + {\cal B} u + \dotsb$, where
\begin{align}
  \mathcal{A} &= -\frac{3}{2} \ii w C_0 + \frac{3}{4} q^2 C_0 +
  \frac{3}{4}\left(C_0 - 3 C_1 \right) M^2\, , \label{eq:xy-calA}\\
  \mathcal{B} &= C_0 + \frac{iw}{4} C_0 \left[6+ \frac{\pi\sqrt{3}
    }{3} - 3\log{3} \right] - \frac{3}{4} C_0 q^2 - \frac{3}{4}\left(
    C_0 - 3 C_1 \right) M^2\, . \label{eq:xy-calB}
\end{align}
The values of $C_0$ and $C_1$ can be fixed in terms of the source
$h_x^{y(0)}$ by recalling the definitions of $\varphi$ and ${\cal A}$
\begin{align}
\varphi (u=0) ={}& h_x^{y(0)} = C_1 \, ,\\
\psi_{xy}(u=0) ={}& \mathcal{A} =r_+^2 (m^2-\omega^2) h_x^{y(0)} =r_+^2
(m^2 - \omega^2) C_1\, .
\end{align} 
Using our expression for ${\cal A}$, \eqref{eq:xy-calA}, we obtain
\begin{equation}
  \label{eq:c1ratio}
  \frac{C_1}{C_0} = \frac{{-}2\ii w + q^2+M^2}{3(2M^2-w^2)}\, .
\end{equation}
Here we've used $4\pi T r_+ \sim 3$, which is valid since to the order
at which we are working. With this ratio in hand, we can readily
calculate the Green's function
\begin{equation}
  \label{eq:Gxyxy-result}
  G_{T^{xy},T^{xy}}^R(\omega,k) = \frac{4\pi T}{3} \frac{2 m^2 - \omega^2}{{-}\ii \omega + \frac{k^2+m^2}{4\pi T}} \, .
\end{equation}
We note that this result does not reproduce the values for $\eta/s$ we
found directly at $k=0$, as we have not worked to high enough
order. Furthermore the $m \to 0$ limit and the $\omega \to 0$ limits
don't commute and to obtain a sensible $\eta/s$, one must first send
$m \to 0$ and then calculate $\eta$, which of course yields $\eta/s =
1/4\pi$. It would be insightful to continue the calculation of
$G^R_{T^{xy},T^{xy}}$ to higher order in $M$, but we were unable to do
so.

\section*{Acknowledgments}
We are grateful to Luca Delacr{\'e}taz and Sean Hartnoll for
insightful discussions. TC was funded for part of the project by the
Stanford Physics Department.

\appendix
\section*{Appendix}

\section{On-shell Action}\label{ap.3}

In order to be self contained, we present in this appendix the terms
in the on-shell action needed to compute the Green's functions found
in the text. As usual, the bulk action must be supplemented by the
Gibbons-Hawking term and the appropriate counter-terms to render the
action finite as the cutoff surface at $r = \epsilon$ is sent to $r=0$
\cite{deHaro:2000vlm, Skenderis:2002wp}. For convenience we reproduce
these terms here:
\begin{equation}
  \label{eq:app-Sgh+ct}
  S_{GH}+ S_{ct} = {-} \frac{1}{2\kappa^2} \int \dd^3 x\, \sqrt{{-}\gamma} \left[ {-} 2 K + R[\gamma] + \frac{4}{L^2} - \frac{1}{2} \sum_i
    \gamma^{ab} \partial_a \phi^i \partial_b \phi^i \right]\, . 
\end{equation}
Here $\gamma$ is the induced metric on the conformal boundary, $K$ is
the trace of the extrinsic curvature of the boundary, and $R[\gamma]$
is the Ricci scalar of the induced metric.

To find the Green's functions, we evaluate the full action, both the
bulk and boundary terms, on solutions to the equations of motion to
quadratic order in the fluctuations. As is typically the case, the
full action actually evaluates to a boundary term; in particular, the
bulk terms can actually be rewritten as a total derivative up to the
equations of motion. To evaluate, we use the near boundary expansions,
\begin{align}
  h^y_t(r) ={}& h^{y(0)}_t + r h^{y(1)}_t + r^2 h^{y(2)}_t  + r^3
  h^{y(3)}_t + \dotsb \, ,\\
  h^y_x(r) ={}& h^{y(0)}_x + r h^{y(1)}_x  + r^2 h^{y(2)}_x  + r^3
  h^{y(3)}_x + \dotsb \, ,\\
  \delta \phi_2(r) ={}& \delta \phi_2^{(0)} + r \delta \phi_2^{(1)} 
  + r^2 \delta \phi_2^{(2)}  + r^3 \delta \phi_2^{(3)} + \dotsb \, ,
\end{align}
and note that the equations of motion automatically only the
coefficients of $r^0$ and $r^3$ are independent, e.g.~$h^{y(1)}_t$ and
$h^{y(2)}_t$ are fixed in terms of $h^{y(0)}_t$ while $h^{y(3)}_t$ is
not. However, this independence is short lived as we also have to
impose the appropriate boundary conditions at the horizon, which
imposes a relation between the two coefficients. Nevertheless, the
dictionary tells us that the leading terms correspond to sources and
we'll see shortly that the ${\cal O}(r^3)$ terms encode the
responses. Since we are only interested in the stress tensor
correlators, we set $\delta \phi_2^{(0)} = 0$.

Carrying through the exercise of plugging these expansions into the
action, we find (up to terms quadratic in the sources, which
correspond to contact terms we ignore)
\begin{align}
  S_{\text{on-shell}} = \frac{3}{4\kappa^2} \int \frac{\dd \omega \dd
    k}{(2\pi)^2} \left[ {-} h^{y(0)}_t({-}\omega,{-}k)
    h^{y(3)}_t(\omega,k) + h^{y(0)}_x({-}\omega,k)
    h^{y(3)}_x(\omega,k) \right]\, .
\end{align}
Here we see explicitly that the response, obtained by the derivative
of the action with respect to the source, is directly given by
$h^{y(3)}_t$ and $h^{y(3)}_x$, and so upon dividing by the source, we
obtain the expressions given for the Green's functions in the text
\begin{align}
  G_{T^{ty},T^{ty}}^R(\omega,k) ={}& \frac{3}{2\kappa^2}
  \frac{h^{y(3)}_t}{h^{y(0)}_t}\, , & G_{T^{xy},T^{xy}}^R(\omega,k)
  ={}& \frac{3}{2\kappa^2} \frac{h^{y(3)}_x}{h^{y(0)}_x}\, .
\end{align}

While the expansions given above, namely those of the precise metric
and scalar fluctuations, are physically transparent, we saw in the
text that decoupling the equations of motion was aided by defining
combinations of the bare fields, e.g.~$\psi_{ty}$ and $\psi_{xy}$. It
is a simple matter, discussed in the text, to use the definitions of
these gauge invariant combinations to relate their near boundary
expansions to the near boundary expansions of $h^{y}_t$ and $h^y_x$.

\section{$G_{T^{ty}T^{ty}}$ fluctuation equations}\label{ap.1}
Plugging the perturbative expansion \eqref{eq:psi-ty-pert} into
\eqref{eq:psi-ty-tilde-eq}, one obtains the following equations
\begin{align}
  0 ={}& \left[(1-u^3) \partial_u^2 - 3
    u^2 \partial_u  + 3u\right] \tilde \psi_{ty}^{(0)}\, , \\
  0 ={}& \left[(1-u^3) \partial_u^2 - 3 u^2 \partial_u + 3u\right]
  \tilde \psi_{ty}^{(1)} + 3u(u\partial_u + 1) \tilde \psi_{ty}^{(0)}
  \, , \\
  0 ={}& \left[(1-u^3) \partial_u^2 - 3 u^2 \partial_u + 3u\right]
  \tilde \psi_{ty}^{(2)} - \frac{9}{4} \tilde \psi_{ty}^{(0)}
  \, , \\
  0 ={}& \left[(1-u^3) \partial_u^2 - 3 u^2 \partial_u + 3u\right]
  \tilde \psi_{ty}^{(3)} - \frac{9}{8} u\left[u(1-u) \partial_u^2
    +(2-3u)\partial_u + 3 \right]\tilde \psi_{ty}^{(0)} \, .
\end{align}
Each of these can be readily solved. Requiring the solutions to be
regular at the horizon and $\tilde \psi_{ty}^{(i)}(1) = 0$ for $i>0$,
we find the following solutions
\begin{align}
  \tilde \psi_{ty}^{(0)}(u) ={}& C u\, , \label{eq:apA-psi0-sol}\\
  \tilde \psi_{ty}^{(1)}(u) ={}& {-}\frac{C}{12} \left\{18 + u \left[
      9 \log \left(\frac{3}{1+u+u^2}\right) + 6 \sqrt{3} \tan^{-1}
      \left(\frac{1+2u}{\sqrt{3}}\right) - 18 - 2 \sqrt{3} \pi \right]
  \right\}\, , \label{eq:apA-psi1-sol}\\
  \tilde \psi_{ty}^{(2)}(u) ={}& \tilde \psi_{ty}^{(3)}(u) =
  \frac{3C}{4} (1-u)\, . \label{eq:apA-psi23-sol}
\end{align}
These are all the terms needed for the solution given in
\eqref{eq:psi-tysol}.

To obtain the order $M^2$ contribution to the dispersion relation, we
need to solve the higher order terms in the expansion
\eqref{eq:psi-ty-pert}. These terms are governed by the following
equations
\begin{align}
  0 ={}& \left[(1-u^3) \partial_u^2 - 3 u^2 \partial_u + 3u\right]
  \tilde \psi_{ty}^{(4)} + \frac{9}{8} \left[u (2 - 3 u) \partial_u +
    (1-3u) \right] \tilde \psi_{ty}^{(0)} \\
  &+ \frac{9}{8} u \left[ u (1-u) \partial_u^2 + (2-3u) \partial_u +3
  \right] \tilde \psi_{ty}^{(1)} + 3u (u\partial_u + 1) \tilde
  \psi_{ty}^{(3)} \nonumber \\
  0 ={}& \left[(1-u^3) \partial_u^2 - 3 u^2 \partial_u + 3u\right]
  \tilde \psi_{ty}^{(5)} + \frac{9}{16} \left(4 \tilde \psi_{ty}^{(3)}
    - 3 \tilde \psi_{ty}^{(0)} \right) \\
  &+ \frac{9}{8}u \left[u(1-u) \partial_u^2 +(2-3u) \partial_u + 3
  \right] \tilde \psi_{ty}^{(2)} \, .\nonumber
\end{align}
Using the solutions,
\eqref{eq:apA-psi0-sol}--\eqref{eq:apA-psi23-sol}, for $\tilde
\psi_{ty}^{(i)}$ for $i<4$, we can solve these, again imposing
regularity at the horizon and $\tilde \psi_{ty}^{(4,5)} = 0$, to find
\begin{align}
  C^{{-}1}\tilde \psi_{ty}^{(4)}(u) ={}& \frac{\sqrt{3}}{8}
  (1+u)\left[\pi - 3 \tan^{{-}1} \left( \frac{1+2u}{\sqrt{3}}\right)
  \right] \nonumber \\
  &+ \frac{9}{16(1+u+u^2)} \left[1 + u^2 - 2 u^3 - (1-u^3) \log
    \left(\frac{3}{1+u+u^2}\right) \right] \, ,  \\
  C^{{-}1}\tilde \psi_{ty}^{(5)}(u) ={}& {-} \frac{3\sqrt{3}}{16} u
  \left[ \pi - 3 \tan^{{-1}} \left(\frac{1+2u}{\sqrt{3}} \right)
  \right]\, .
\end{align}
Series expanding these solutions near the boundary leads to the order
$w M^2$ and $q^2 M^2$ terms in \eqref{eq:ty-A-m2}. 

\subsection{Non-perturbative hydrodynamic terms}
As mentioned in the text, the dispersion relation can actually be
solved for exactly in $M$. The solution itself is very lengthy and not
worth producing directly here, so we will simply present the equations
needed to obtain the solution and leave it to the reader to plug the
equations below into \texttt{Mathematica}.

To set up the problem, we expand $\tilde \psi_{ty}$ as before but
letting the functions depend on $M$
\begin{align}
  \tilde \psi_{ty}(u) = \tilde \psi_{ty}^{(0)}(u;M) + \ii w \tilde
  \psi_{ty}^{(1)}(u;M) + q^2 \tilde \psi_{ty}^{(2)}(u;M) + \dotsb\,
  . \label{eq:apA-non-pert-exp}
\end{align}
Since we aren't working perturbatively in $M$, we don't need to expand
$R_+$ in $M$, and therefore, we can work directly in terms of the
emblackening factor. Then the coupled equations governing the
expansion \eqref{eq:apA-non-pert-exp} are
\begin{align}
  0 ={}& \partial_u \left[f \partial_u\tilde \psi_{ty}^{(0)} \right] +
  \frac{3}{2u^2} \left(2 - M^2 R^2 u^2 - 2 f \right) \tilde
  \psi_{ty}^{(0)}\, , \\
  0 ={}& \partial_u \left[f \partial_u\tilde \psi_{ty}^{(0)} \right] +
  \frac{3}{2u^2} \left(2 - M^2 R^2 u^2 - 2 f \right) \tilde
  \psi_{ty}^{(1)} - \partial_uf \partial_u \tilde \psi_{ty}^{(0)} -
  \frac{1}{2} \tilde \psi_{ty}^{(0)} \partial_u^2 f \, ,
  \\
  0 ={}& \partial_u \left[ f \partial_u \tilde \psi_{ty}^{(2)} \right]
  + \frac{3}{2u^2} \left(2 - M^2 R^2 u^2 - 2 f \right) \tilde
  \psi_{ty}^{(2)} - R^2 \tilde \psi_{ty}^{(0)}\, .
\end{align}
These equations can, with a bit of patience, can be solved upon using
the expression for the emblackening factor, imposing regularity at the
horizon and $\tilde \psi_{ty}^{(1)}(1) = \tilde \psi_{ty}^{(2)}(1) =
0$. The solutions obtained in this fashion can be readily expanded for
small $M$ to check against the perturbative expressions given
previously.

\section{Solving $\varphi (r)$}\label{ap.2}

By combining equations \eqref{eq:txy-hxy-eq-k0} and
\eqref{eq:psi-xy-pert}, we deduce that
\begin{equation}
h_x^{y\prime\prime } + \left[\frac{f'}{f}-\frac{2}{u} \right] h_x^{y\prime} + \frac{R_+^2(w^2-q^2 f-M^2 f )}{f^2} h_x^y = 0
\end{equation}
Plugging in the ansatz $h_x^y(u) = f(u)^{-i \frac{w}{2}} F(u)$:
\begin{equation*}\label{phiFODE}
  0 = F'' + \left[(1-\ii w)\frac{f'}{f} - \frac{2}{u} \right] F' + \left[ \ii \frac{w}{u} \frac{f'}{f} - \ii \frac{w}{2} \frac{f''}{f}+ \frac{w^2}{f^2} \left(R_+^2 - \frac{f'^2}{4}\right) - \frac{R_+^2(q^2+M^2)}{f} \right] F.
\end{equation*}
We can thus solve $F(u)$ perturbatively:
\begin{equation}
F(u)=F_0(u)+wF_1(u)+q^2G_1(u)+w^2 F_2(u)+x^2 H_1(u) + wx^2 J1(u) + x^4 H_2(u) + \dots
\end{equation}
We obtain 4 equations:
\begin{align}
  0={}&\left[(1-u^3)\partial_u^2 -\left(\frac{2+u^3}{u} \right)  \partial_u\right]  F_0\, , \\
  0={}& \left[(1-u^3)\partial_u^2 -\left(\frac{2+u^3}{u} \right)  \partial_u\right]  F_1 +3 \ii u^2 \partial_u F_0 \, ,\\
  0={}& \left[(1-u^3)\partial_u^2 -\left(\frac{2+u^3}{u} \right)  \partial_u\right]  G_1   - \frac{9}{4} F_0 \, ,\\
  0={}& \left[(1-u^3)\partial_u^2 -\left(\frac{2+u^3}{u}
    \right) \partial_u\right] H_1 -
  \frac{u(u+2)(1-u)}{2(1+u+u^2)} \partial_u F_0 - F_0 \, .
\end{align}
Each of these can be readily solved. Requiring the solutions to be
regular at the horizon and $F(1)-F_0(1)= 0$,
we find the following solutions
\begin{align}
  F_0(u)={}& C\, ,\\
  F_1(u)={}& 0\, ,\\
  H_1(u)={}& -\frac{C}{18} \left[2 \pi \sqrt{3} -9\log{12}  -6\sqrt{3} \arctan{\frac{1+2u}{\sqrt{3}}} + 9 \log{[-i + \sqrt{3} - 2 iu]} + \right. \\
  &\left. \qquad \quad\vph{\frac{1+2u}{\sqrt{3}}} + 9 \log{[\ii + \sqrt{3}
      + 2\ii u]}\right]\, ,
\end{align}
together with the following equations for  $J_1(u), F_2(u)$ and $H_2(u)$:
\begin{align}
  0={}& J_1''-\left[\frac{3 u^2}{1-u^3}+\frac{2}{u}\right] J_1' - \ii
  C\frac{1+u+u^2+6u^3}{2 \left(1-u^3\right) \left(1+u+u^2\right)} \, ,
  \\
  0={}& F_2''-\left[\frac{3 u^2}{1-u^3}+\frac{2}{u}\right] F_2' +
  \frac{9C}{4} \frac{(1+u)(1+u^2)}{(1-u) \left(1+u+u^2\right)^2} \, ,\\
  0={}& H_2''-\left[\frac{3 u^2}{1-u^3}+\frac{2}{u}\right] H_2'
  -\frac{C}{18} \left[ \frac{9 u^2 (2 u^2 +2u -1)-2 \sqrt{3} \pi
      (1+u+u^2)^2}{(1-u) \left(1+u+u^2\right)^3} + \right. \\
  & \qquad \qquad \qquad \qquad \qquad\left. + \frac{6 \sqrt{3}
      \arctan\left(\frac{2 u+1}{\sqrt{3}}\right)+9 \log{3} -9 \log
      \left(1+u+u^2\right)}{1-u^3} \right] \, .
\end{align}

\bibliographystyle{utphys}
	\bibliography{refs}
	
\end{document}